\documentclass{IEEEtran}
\usepackage[T1]{fontenc}
\usepackage{amsmath}
\usepackage{graphicx}

\makeatletter
 \newcommand{\lyxaddress}[1]{
   \par {\raggedright #1 
   \vspace{1.4em}
   \noindent\par}
 }

\makeatother
\begin{document}

\title{Delayed Correlations in Inter-Domain Network Traffic}

\author{Viktoria Rojkova, Mehmed Kantardzic}

\maketitle

\lyxaddress{Department of Computer Engineering and Computer Science, University
of Louisville, Louisville, KY 40292 email: \{vbrozh01, mmkant01\}@louisville.edu }

\begin{abstract}
To observe the evolution of network traffic correlations we analyze
the eigenvalue spectra and eigenvectors statistics of delayed correlation
matrices of network traffic counts time series. Delayed correlation
matrix $D\left(\tau\right)$ is composed of the correlations between
one variable in the multivariable time series and another at a time
delay $\tau$. We determined, that inverse participation ratio (IPR)
of eigenvectors of $D\left(\tau\right)$ deviates substantially from
that of eigenvectors of the equal time correlation matrix. The largest
eigenvalue $\lambda_{max}$ of $D\left(\tau\right)$ and the corresponding
IPR oscillate with two characteristic periods of $3\tau$ and $6\tau$.
Found delayed correlations between network time series fits well into
the long range dependence (LRD) property of the network traffic.

Injecting the random traffic counts between non-randomly correlated
time series, we were able to break the picture of periodicity of $\lambda_{max}$.
In addition, we investigated influence of the periodic injections
on both largest eigenvalue and the IPR, and addressed relevance of
these indicators for the LRD and self-similarity of the network traffic. 
\end{abstract}

\section*{Categories and Subject Descriptors}

C.2.3 {[}\textbf{Computer-Communication Networks}{]}: Network Operations

\section*{General Terms}

Measurement, Experimentation

\begin{keywords}
Network-Wide Traffic Analysis, Random Correlations, Time-Lagged Correlations,
Long Range Dependence
\end{keywords}

\section{introduction}

The cross-correlation matrix is one of the tools of time series analysis,
used in studies of the underlying interactions between network structural
constituents. The network traffic system is often seen as a collection
of times series with traffic bytes per time interval or number of
packets per time interval at a single link. Within this collection,
temporal correlations between network structural constituents and
consequently patterns of collective behavior might be present. 

To study such patterns one can employ the equal-time cross-correlation
matrix of traffic time series at all network links (see, for example
\cite{Rojkova,Barthelemy}). Further statistical description of the
awareness is commonly done via eigen-decomposition of cross-correlation
matrix. Detailed studies showed that the major portion of eigenvalues
of cross-correlation matrix fall into the theoretically predicted
boundaries of eigenvalues spectrum of random matrices \cite{Rojkova,Barthelemy}.
This portion satisfies the so called universal properties described
by the random matrix theory (RMT) \cite{Guhr3}. It also recognizes
the uncongested state of the traffic, where each router is able to
communicate with any other router and with each subnet under its service.
The part of the spectrum deviating from the RMT boundaries represents
stable in time non-random correlations between network traffic time
series \cite{Rojkova,Barthelemy}.

Even though, such distinction provides a valuable insight into the
meaning of empirical data, the equal-time cross-correlation matrix
on its own may not be sufficient for understanding of the effect of
inter-domain correlations at different times. In fact, the analysis
of correlation as a function of time lags has already been used in
econometric time series systems. For example, the analysis of stock
returns portfolio showed the asymmetric lead-lag relationship between
stock returns; high-volume stocks lead the low-volume stocks \cite{Lo,Chordia}.
This finding is attributed to information adjustment asymmetry \cite{Biely}.
The uncongested balanced state of the network traffic implies the
\char`\"{}symmetric\char`\"{} information/traffic flow exchange. Thus,
to control the congestion level of network traffic systems, it is
crucial to recognize the collective behavior or correlation patterns
between network traffic time series, and to observe their evolution
in time. 

In order to trace the evolution of correlation pattern between traffic
time series the equal-time cross-correlation matrix is replaced with
time-lagged correlation matrix \cite{GlupyeIndusy}. The obtained
matrix of delayed correlations $D\left(\tau\right)$ can be subjected
to eigen-analysis just like its equal-time counterpart. In contrast
to equal-time correlation matrices which have a real eigenvalue spectrum,
the spectrum of $D\left(\tau\right)$ is complex since matrices of
these types are asymmetric. While the general properties of complex
spectrum of $D\left(\tau\right)$ is unknown so far, results of symmetrized
version of lagged correlation matrices have been reported recently
\cite{GlupyeIndusy,Burda}.

In this paper, we concentrated on time-lagged correlations between
time series generated by Simple Network Manage Protocol (SNMP) traffic
counters of University of Louisville backbone routers system. We established
that time-lagged correlations between traffic time series sustain
for up to $100\tau$, where $\tau=300$ sec. Moreover, the largest
and smallest eigenvalues $\lambda_{max}$ and $\lambda_{min}$ of
$D\left(\tau\right)$ and the inverse participation ratio (IPR) of
the eigenvector corresponding to $\lambda_{max}$ were found to oscillate
with two characteristic periods of $3\tau$ and $6\tau$. 

In addition, within the content of the eigenvector corresponding to
$\lambda_{max}$ we determined the index of a single time series which
is driving the $\lambda_{max}$ oscillation. The IPR of all eigenvectors
of $D\left(\tau\right)$ shows the single time series contributor
in the eigenvector which at the equal-time correlations belonged to
the random part of the spectrum. Since the theoretical prediction
for the spectrum of time-lagged correlation matrices is unknown, we
keep the terminology of the spectrum derived from eigen-decomposition
of equal-time correlation matrices. Thus, \emph{random} in our text
refers to the part of the spectrum within the boundaries predicted
by the RMT for equal-time correlation matrix. \emph{Non-random} denotes
the part of the spectrum outside of the RMT predictions. 

Finally, we found that the injection of random traffic counts between
time series which interact \emph{non-randomly} destroys the oscillatory
picture of $\lambda_{max}$ and $\lambda_{min}$ and that of the corresponding
IPRs. Meanwhile, the injection of traffic counts with smooth (periodic)
time dependence between \emph{randomly} interacting traffic time series
with respect to characteristic periods of $\lambda_{max}$, $\lambda_{min}$
and IPR reveals new periodicity (of $4\tau$). The findings of eigenvalue
spectrum analysis and experimental results suggest the asymmetric
long lasting relationship between traffic time series.

Network traffic analysis had undergone the evolution from considering
the network traffic time series as an outcome of Poisson and memory-less
processes to recognizing the long range dependencies and self-similarity
of the traffic. The statistics of eigenvalue spectrum and IPR of eigenvectors
of time-lagged correlation matrices provide essential dimensional
reduction in the investigation of long-ranged dependence (LRD) of
the network traffic. Before starting the expensive process of Hurst
parameter estimation, one can attempt to find just a few indicators
of LRD, self-similarity or deviations from thereof.

The rest of the paper is organized as follows. We introduce time-lagged
correlation matrix and its eigensystem in Section II. Section III
is devoted to numerical analysis of time-lagged eigensystem. Then,
in Section IV by experimenting with the content of the original time
series we addressed LRD and self-similarity of lagged correlations
through proposed indicators. Conclusions and discussion are given
in Section V.

\section{time-lagged correlation matrix of network traffic time series}

The starting point of our discussion is averaged traffic count data
collected from router-router and router-VLAN subnet connections of
the University of Louisville backbone routers system. The same data
set was used in \cite{Rojkova} in the context of equal-time correlation
matrix analysis. Below we recap the details relevant to the construction
of lagged correlation matrix, relegating the information on the network
and the way data was processed to \cite{Rojkova}. We will be dealing
with total of $L=2015$ records of $N=497$ time series averaged over
$300$ seconds, where incoming and outgoing traffic generate independent
time series.

In order to define traffic rate change $G_{i}\left(t\right)$ we used
the logarithm of the ratio of two successive counts upon calculating
the traffic rate change of time series $T_{i}$, $i=1,\dots,N$ ,
over time $\Delta t$,\[
G_{i}\left(t\right)\equiv\textrm{ln}\, T_{i}\left(t+\Delta t\right)-\textrm{ln}\, T_{i}\left(t\right),\]
 \cite{Barthelemy}. Then we introduce normalization, according to
$g_{i}\left(t\right)=\left(G_{i}\left(t\right)-\left\langle G_{i}\left(t\right)\right\rangle \right)/\sqrt{\left\langle G_{i}^{2}\right\rangle -\left\langle G_{i}\right\rangle ^{2}}$
and built the time-lagged correlation matrix $D\left(\tau\right)$
as follows \cite{GlupyeIndusy}\begin{align}
D_{ij}\left(\tau\right) & \equiv\textrm{{Sym}}\left\langle g_{i}\left(t\right)g_{j}\left(t+\tau\right)\right\rangle \nonumber \\
 & =\frac{1}{2L}\sum_{t=0}^{t=L}\left(g_{i}\left(t\right)g_{j}\left(t+\tau\right)+g_{j}\left(t\right)g_{i}\left(t+\tau\right)\right).\label{DeeOfTau}\end{align}
Here the sole purpose of symmetrization is the restriction of the
eigenvalues and eigenvectors to real values. In principle, the numerical
experiments we run below can be repeated for the eigensystem of non-symmetric
correlation matrix. Studies of the latter are already in progress
in a different setting (see, for example, \cite{Biely}).

\subsection{Eigenvalues and eigenvectors}

Next we proceed with defining the indicators we focus on in what follows.
First of all, the eigenproblem for our cross-correlation matrix is
time dependent\begin{equation}
D\left(\tau\right)u_{k}\left(\tau\right)=\lambda_{k}\left(\tau\right)u_{k}\left(\tau\right),\label{EigenProblem}\end{equation}
where $\lambda_{k}$ is $k$-th eigenvalue, corresponding to $k$th
eigenvector. In other words, the eigensystem $\left\{ \lambda_{k}\left(\tau\right),\, u_{k}\left(\tau\right)\right\} $
is defined for each increment of delay time. As opposed to same time
eigensystem $\left\{ \lambda_{k}\left(0\right),\, u_{k}\left(0\right)\right\} $,
our eigensystem does not characterize presence or lack of organization
(localization) in the system at a given time. Instead it can serve
as a measure of back-in-time (or forward in time, depending on prospective)
correlation within the network structure.

Furthermore, the RMT picture of the eigenvalues and eigenvectors,
in which spectrum is split into three parts is no longer valid. By
three parts we understand the RMT part, which behave universally at
the center of the eigenvalue spectrum, and {}``left'' and {}``right''
parts (lying to the left and to the right from predicted RMT bounds)
which exhibit non-universal features \cite{Rojkova}. Although, for
very small $\tau$ this subdivision is clearly still accurate, we
can expect, transient behavior of $\left\{ \lambda_{k}\left(\tau\right),\, u_{k}\left(\tau\right)\right\} $
to reveal new, otherwise undetectable correlations within the network.
Hence, we found it convenient to keep track of quantitative and qualitative
changes in eigensystem using left/random/right terminological distinction.

On the other hand, observing the entire system of $497$ eigenvalues
and the same amount of eigenvectors can be quite tedious task. One
needs to come up with more concise indicators of network behavior.
Selecting efficient indicators can help in defining {}``normal''
state of the system, a task quite challenging on its own, and in predicting
structural reaction to the external or internal disruption. Indicators
can be chosen based on the experiments described below, or on their
advanced variations. The candidates are those eigenvalues, that are
most receptive to a particular probe. As far as eigenvectors are concerned
we decided to test corresponding IPRs.

\subsection{Inverse participation ratio}

Given the eigenvector $u_{k}\left(\tau\right)$ the IPR is computed
according to\begin{equation}
I_{k}\left(\tau\right)\equiv\sum_{l=1}^{N}\left[u_{k}^{l}\left(\tau\right)\right]^{4},\label{eq18}\end{equation}
with $u_{k}^{l}$, $l=1,\dots,497$ stands for components of the $k$th
eigenvector \cite{Guhr3}. The IPR is quite indicative in terms of
signaling the number of significant $u_{k}^{l}$, i.e. {}``contributors''
to the eigenvector of interest. For example, if we have reasons to
expect absence of correlations between routers input into the experimental
data, $I_{k}\left(0\right)$ should have its value around $1/\sqrt{N}$.
Indeed, the eigenvector is normalized, thus $\sum_{l=1}^{N}\left[u_{k}^{l}\left(\tau\right)\right]^{2}=1$.
It has $N$ components, and they are all roughly the same in magnitude
(otherwise correlations must be present). Therefore, $u_{k}^{l}\simeq1/\sqrt{N}$,
and $I_{k}\left(0\right)\simeq1/N$. Note, that since $N$ is typically
much greater than $1$, any finite value of IPR signals \emph{localization}
in inter-VLAN traffic \cite{Rojkova}. For non-zero values of $\tau$,
IPRs acquire more general meaning in a sense that routers which interact
heavily at time $t$ may loose their {}``bond'' at time $t+\tau$,
while these not knowing about one another at time $t$ may have significant
mutual contribution at time $t+\tau$. Other more complex possibilities
can be perceived via $I_{k}\left(\tau\right)$ as well.

\section{Eigenanalysis for time-lagged correlation matrix}

\subsection{Stroboscopic sequence for eigensystem}

Upon building the cross-correlation matrix $D\left(\tau\right)$ with
the help of Eq. (\ref{DeeOfTau}) we perform eigen-decomposition (see
Eq. (\ref{EigenProblem})) numerically, using standard \emph{MATLAB}
routine. We look at the result of calculation of eigenvalues for all
delay times $\tau$. A noticeable spike for very small values of delay
time is expected, notwithstanding the position in spectrum. However,
our increments in $\tau$ ($=300\, sec$) may not be small enough
to observe it. For the remainder of observation the result has to
uncover the way system constituents communicate with themselves and
their neighbors on a long run.%
\begin{figure*}
\begin{center}\includegraphics{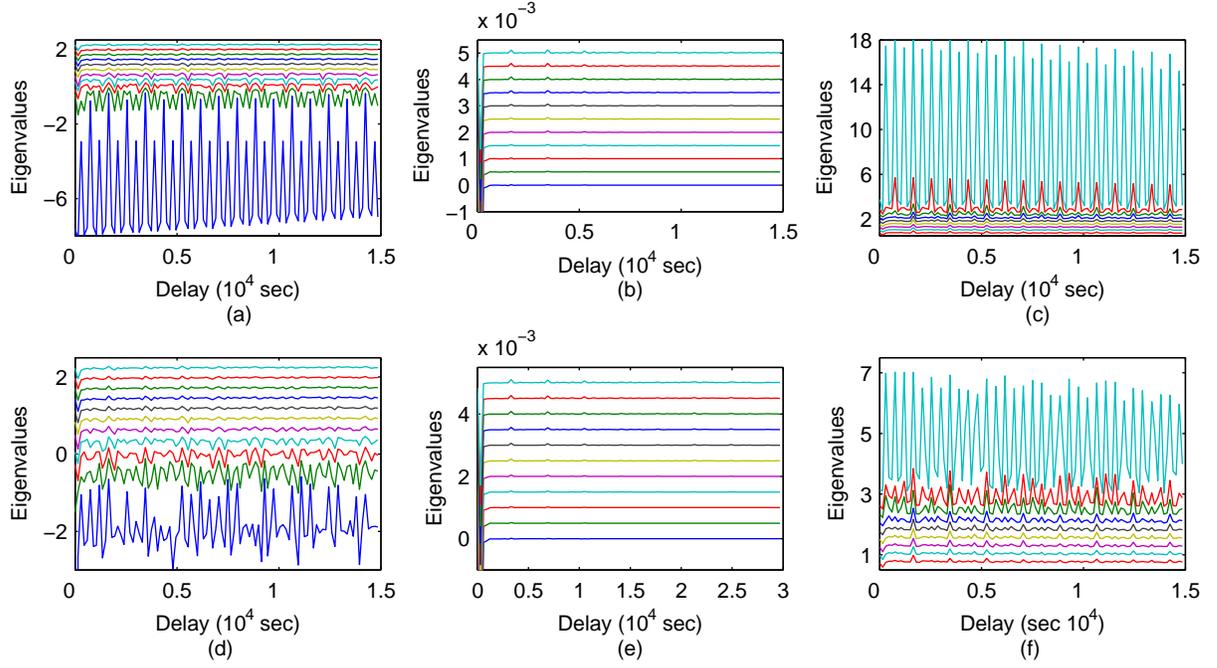}\end{center}

\caption{\label{1} (a) left, (b) random, and (c) right parts of the eigenvalue
spectrum as obtained from actual data. Same graphs are presented in
(d), (e), (f) respectively, after noise-like injections are made. }
\end{figure*}
\begin{figure*}[t]
\begin{center}\includegraphics{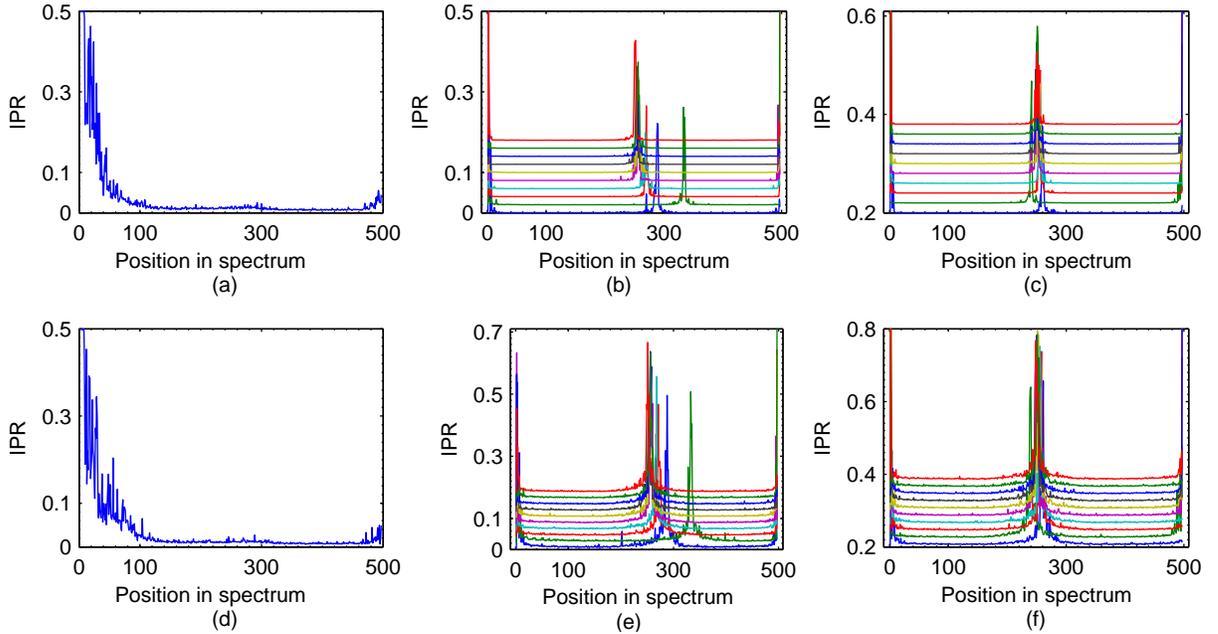}\end{center}

\caption{\label{2} (a) $I\left(0\right)$ versus position in spectrum. Stroboscopic
representation of IPRs corresponding to (b) first $10\tau$ (c) second
$10\tau$; (d)-(f) are the same representations upon noise-like injections.}
\end{figure*}
In Fig. \ref{1}(a)-(c) we illustrate how left/random/right structure
of the spectrum evolves with $\tau$. As it turns out, {}``randomness''
and {}``regularity'' find their new interpretations in the context
of system remembering itself after $\tau$ has elapsed. With the exception
of a few located at the right and left edges of the spectrum, most
eigenvalues fall very close to each other numerically. To make it
more transparent we plotted their $\tau$-dependence using different
offset values (these values are the same within each part). Only ten
eigenvalues are offseted in each case and plotted versus time. The
lowest eigenvalue was excluded from consideration here and throughout
the paper due to it secular behavior in $\tau$.

At a glance, non-edge eigenvalues Fig. \ref{1}(b) with the exception
of an expected spike at small $\tau$ does not seem to represent any
process. Such a lack of forward in time correlation is not completely
surprising, as the eigenvalues from middle part of the spectrum were
referred as RMT-like. It follows, that the random interactions between
traffic time series are time delay invariant. In other words, random
spectrum of eigenvalues is an indicator of self-similarity \cite{Faloutsos}.
Meantime, the eigenvalues at the edges (Figs. \ref{1}(a) and (c))
represent a quasiperiodic process, distinguishing themselves from
their \char`\"{}peers\char`\"{}, the eigenvalues belonging to regular
part of the spectrum for $\tau=0$ quantitatively, and both qualitatively
and quantitatively from the eigenvalues belonging to the RMT-part
for $\tau=0$. They scale with delay time and clearly exhibit long
time dependence \cite{Faloutsos}. Therefore, it makes sense to look
further into the properties of edge eigenvalues, especially into the
properties of those with relatively high absolute values. The actual
values might be used as a measure of delayed time correlations. Having
located potential indicators we move ahead with the search for others.

A remarkable property of the IPR for equal time cross-correlation
matrix (Fig. \ref{2}(a)) is its consistently low, order $1/N$, value
for the major part of the spectrum. This segment in Fig. \ref{2}(a)
is known to obey the RMT \cite{Rojkova}. To the left and to the right
from this segment there is a strong evidence of regular, non-random
behavior. Now, if the first $20$ instances are considered as in Figs.
\ref{2}(b) and (c), where IPRs offseted by an arbitrary amount for
transparency and plotted versus the eigenvalue position, the situation
looks drastically different. The peak located close to the center
of the spectrum signifies the presence of previously undetected correlations,
and the lead-lag relationship between time series.

Close examination of Figs. \ref{2}(b) and (c) shows, that initially,
the high IPR has changing support in the spectrum. Furthermore, as
explained in Section 2.2, peak value tells us, that about four time
series drive the correlation pattern. Later on, the peak \char`\"{}settles
down\char`\"{} and establishes itself around median eigenvalue position
(Fig. \ref{2}(b)). The meaning of this and other two peaks differs
from that of the IPR peaks in Fig. \ref{2}(a). The increase in IPR
computed from the time delayed matrix $D\left(\tau\right)$ indicate
correlations between system's behavior at a given time and its stroboscopic
image after $\tau$, rather than correlations within the spectrum.
In addition, it provides reasonable way of tracking down the sources
of lead-lag behavior. Thus, the observed features make IPR a good
candidate for an indicator of the network congestion state. Note also
significant change in height of the central peak.

\subsection{Frequency domain analysis}

\begin{figure*}[t]
\begin{center}\includegraphics{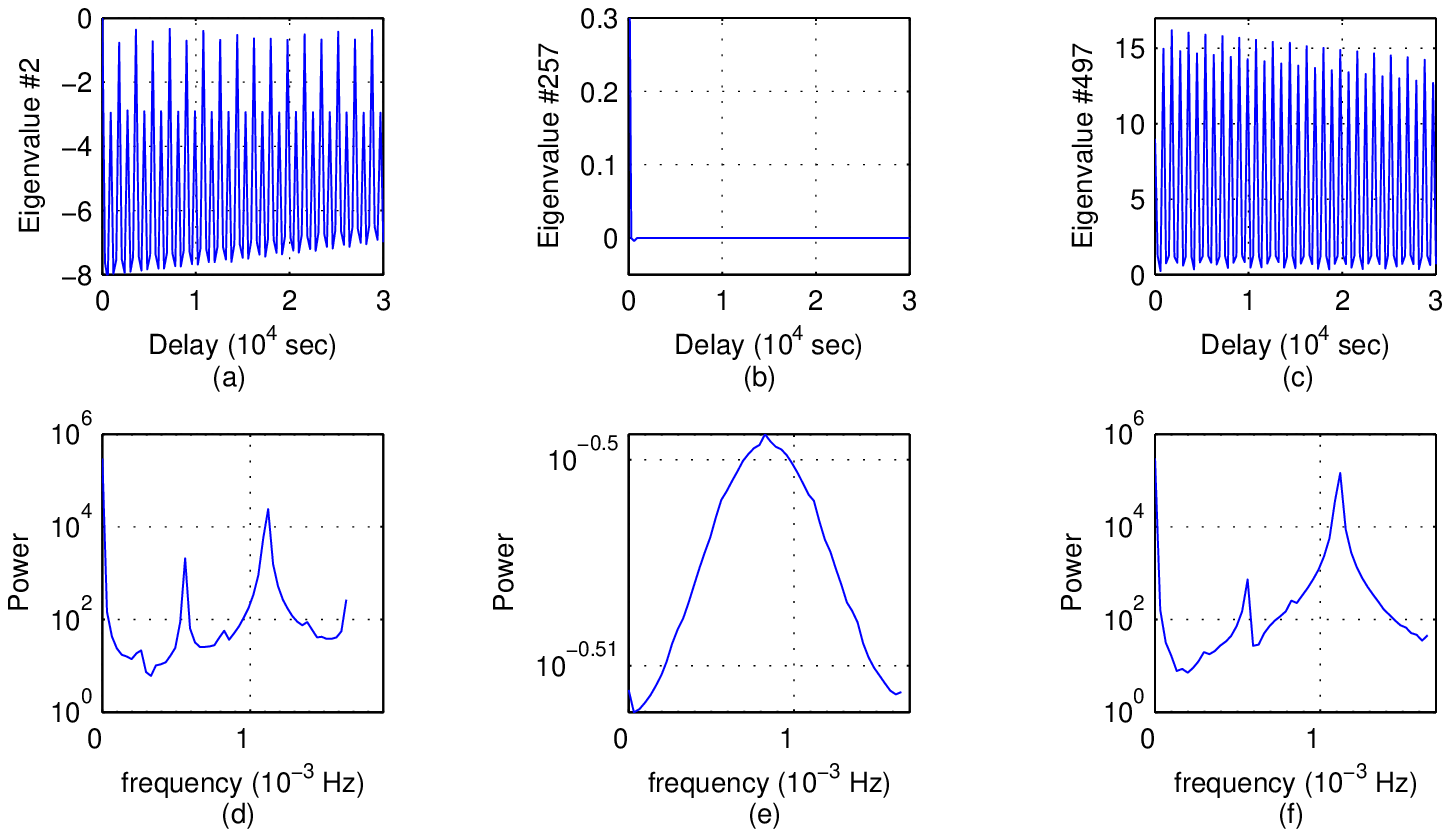}\end{center}

\caption{\label{3} Eigenvalues number (a) $2$, (b) $257$, and (c) $497$,
plotted with respect to time and their respective Fourier spectra
((d) through (f)). }
\end{figure*}
\begin{figure*}
\begin{center}\includegraphics{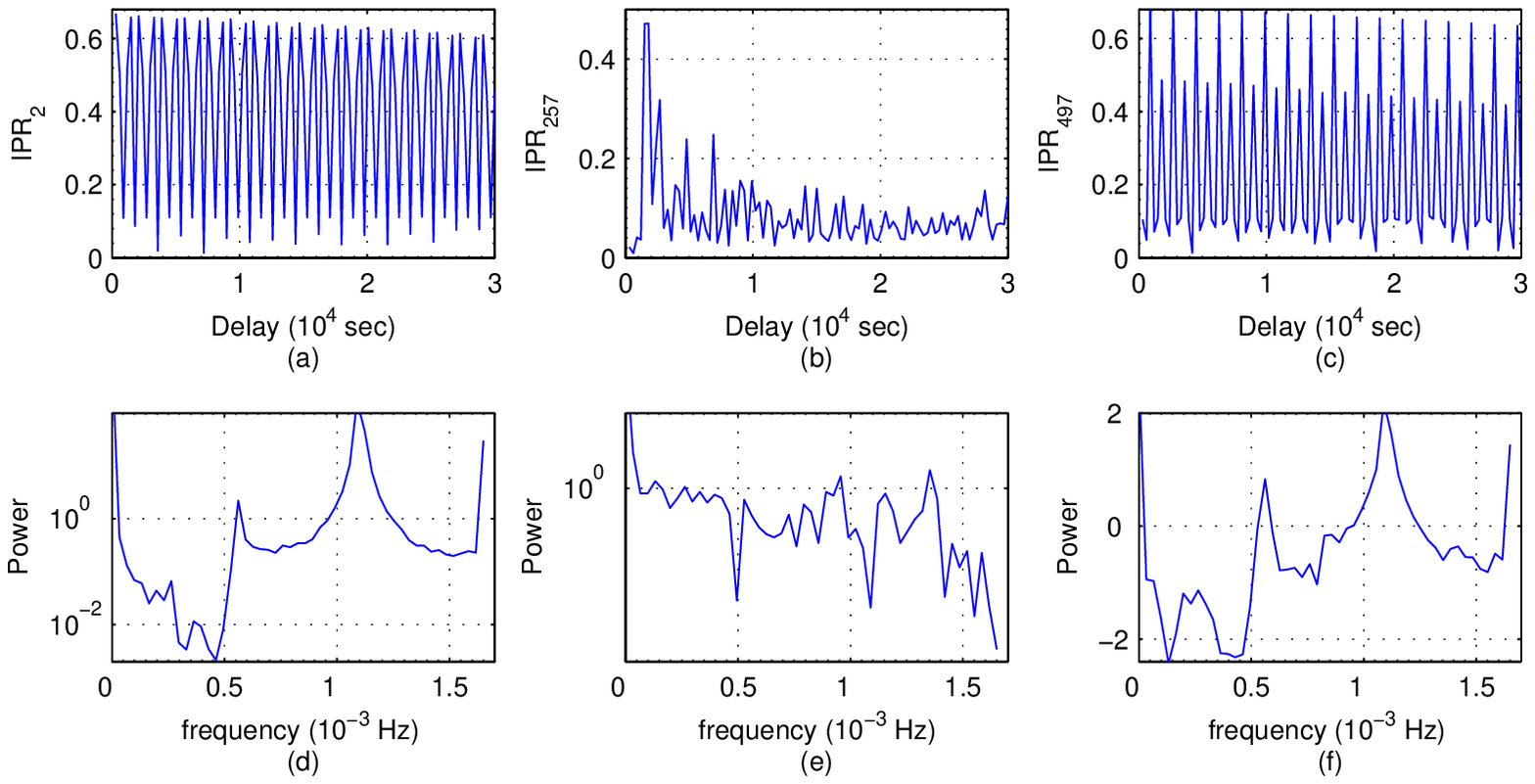}\end{center}

\caption{\label{4} IPRs for eigenvalues number (a) $2$, (b) $257$, and
(c) $497$, plotted with respect to time and their respective Fourier
spectra ((d) through (f)). }
\end{figure*}

The next step in getting more quantitative on long memory processes
in network traffic is to analyze transient behavior of eigenvalues
and IPRs of matrix $D\left(\tau\right)$ in detail. Since quasiperiodic
behavior is present in the majority of quantities of interest we focus
on their frequency content. The standard way to proceed is to transform
$\lambda_{i}\left(\tau\right)$ into frequency domain using fast Fourier
transform. In a sense, we construct a spectra of the spectrum. The
same operation is performed on respective IPRs. We take fast Fourier
transforms for all of the functions at hand, and then, take the square
of their absolute value. The result is referred to as a power spectrum.
There should be no confusion, as graphs of power always accompany
the corresponding time domain quantity. 

In Fig. 3 we display representative eigenvalue dynamics. Once again,
random $\lambda_{i}$ (Fig. 3(b)) does not exhibit anything remarkable,
compared to its regular counterparts. The latter resemble each other,
reflecting a symmetry of the spectra induced by symmetrizing procedure
(Eq. (\ref{DeeOfTau})). For now, we can talk about them in parallel.

Aside from a substantial low and high frequency contribution, which
could have already been guessed from Figs. 3(a) and (c), we discover
two strong contributions from frequencies corresponding to oscillations
with time periods $15$ and $30$ minutes respectively (cf. Figs.
3(d) and (f)). This is in evident contrast to the situation with power
of a random eigenvalue. Such an eigenvalue has equal (negligibly small)
contribution from the entire range of frequencies. The existence of
these two characteristic frequencies suggests a natural way assessing
the current state of the inter-domain network traffic. In fact, it
might be possible to use these as the LRD quantifier estimators \cite{Faloutsos}
in the future.

\section{experiments with altering actual network traffic }

\subsection{Noise-like injections}

Next, we investigated consequences of modifying the time-lagged correlations
between time series. We have already known the time series contributing
the most to the correlation pattern \cite{Rojkova}. All of them can
be linked to eigenvalues which fall in what we term here as the right
segment of eigenvalue spectrum. In these series we replaced the original
traffic counts with counts obtained by random number generator for
a certain period of time. Then, we constructed matrix $D\left(\tau\right)$
for all hundred increments and repeated manipulations described in
previous Section 3.1. The results are shown in Fig. \ref{1}(d) through
(f).

The eigenvalues belonging to middle of the spectrum are completely
unaffected, i.e. they are still time delay invariant. Clearly, our
manipulations with the traffic are not disturbing the self-similar
nature of delayed correlations. However, edge eigenvalues loose time
scales present in their original transient behavior (see Fig. \ref{1}
(d) through (f)). In other words, the LRD gets destroyed. Effect on
the IPR (Figs. \ref{2}(d) and (f).) is less noticeable but is still
there, while for the random segment it is absent. The result of random
counts injections can be summarized as presence of randomly positioned
of small peaks superimposed on the original IPR picture. Indeed, in
Figs. \ref{2}(a) and (c) small peaks are very infrequent and unstable
in time, unlike these in Figs. \ref{2}(d) and (f). %
\begin{figure*}
\begin{center}\includegraphics{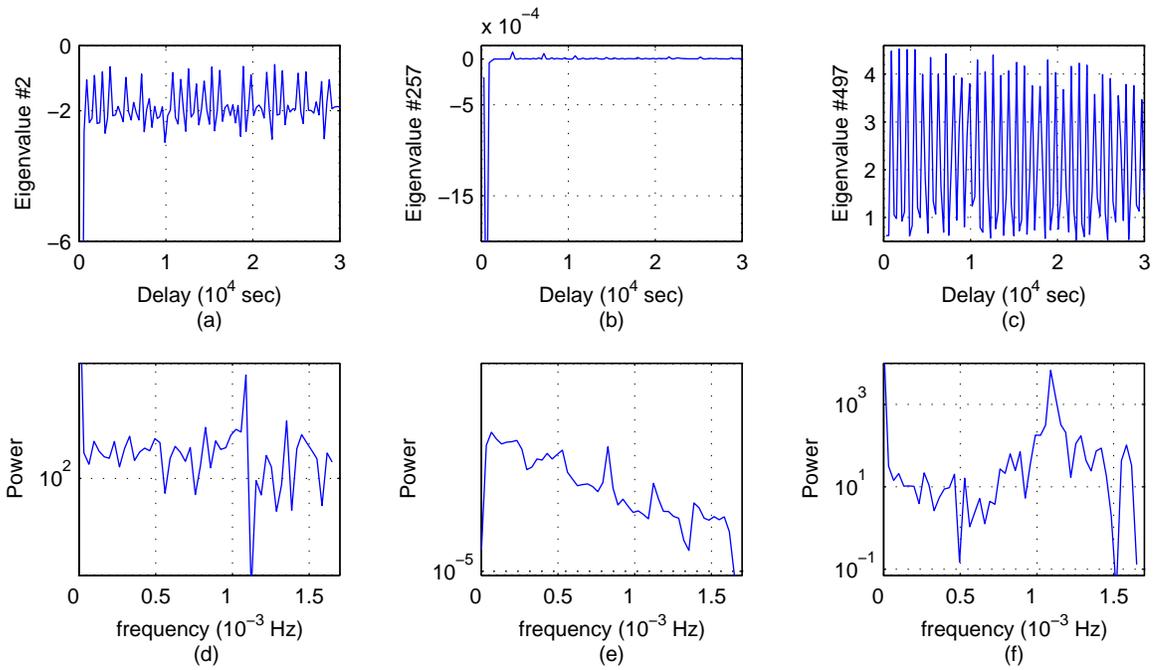}\end{center}

\caption{\label{5} Eigenvalues number (a) $2$, (b) $257$, and (c) $497$,
plotted with respect to time and their respective Fourier spectra
((d) through (f)) after noise-like sample was injected.}
\end{figure*}
\begin{figure*}
\begin{center}\includegraphics{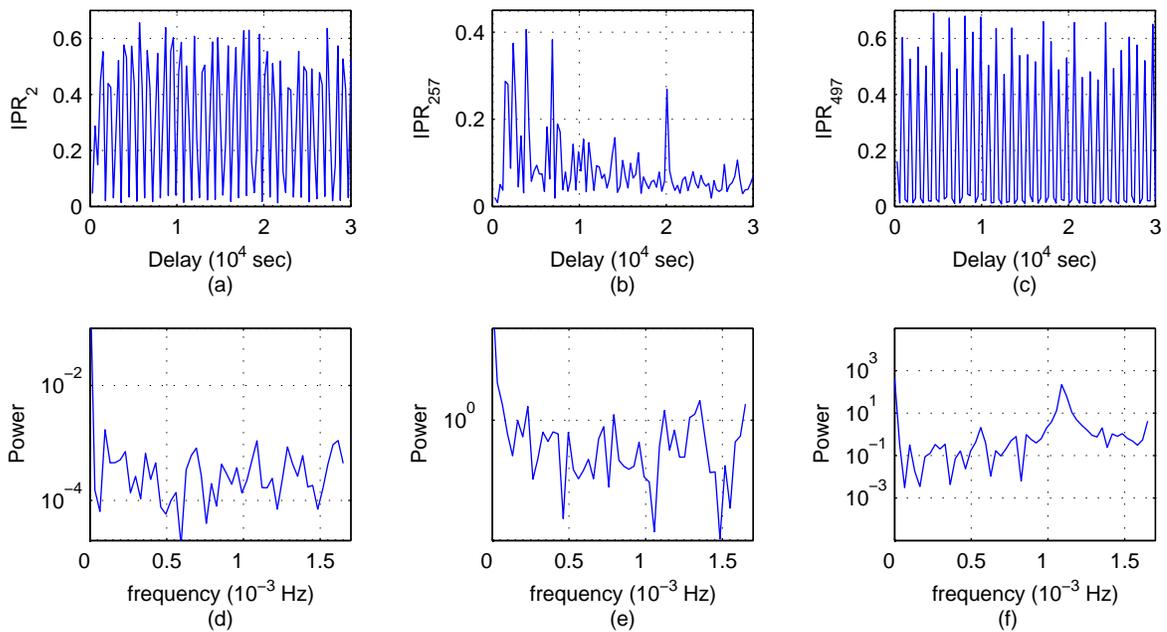}\end{center}

\caption{\label{6} IPRs for eigenvalues number (a) $2$, (b) $257$, and
(c) $497$, plotted with respect to time and their respective Fourier
spectra ((d) through (f)) after noise-like sample was injected.}
\end{figure*}

The above outcome calls for a more close look into eigenvalues and
IPRs for the system experienced the noisy injections into the time
series, which are believed to have major contribution to the overall
traffic in router network. We present three eigenvalues considered
in Section 3.2 as the functions of time delay together with their
respected power spectra. As can be concluded from Fig. 5(a) and (c),
the time dependence looses its LRD structure. It is backed up by the
fact that a lot more frequencies contribute to power spectra upon
random injection. Middle part of the spectra also undergoes certain
transformation, but is still scale-free Fig. 5(b), as actual values
of power are small relative to the power corresponding to edge eigenvalues.
The quantitative changes are also in place for both edge eigenvalues.
The effect can be judged based on comparison of the tallest peaks
in Figs. 5(d) and (f) to their counterparts in Figs. 3(d) and (f).

Similar conclusions can be derived for the IPR as we take a look at
Fig. 6(d) and (f) and compare the outcome of our experiment with the
graphs in Fig. 4(d) and (f).

\subsection{Periodic in time injections}

\begin{figure*}
\begin{center}\includegraphics{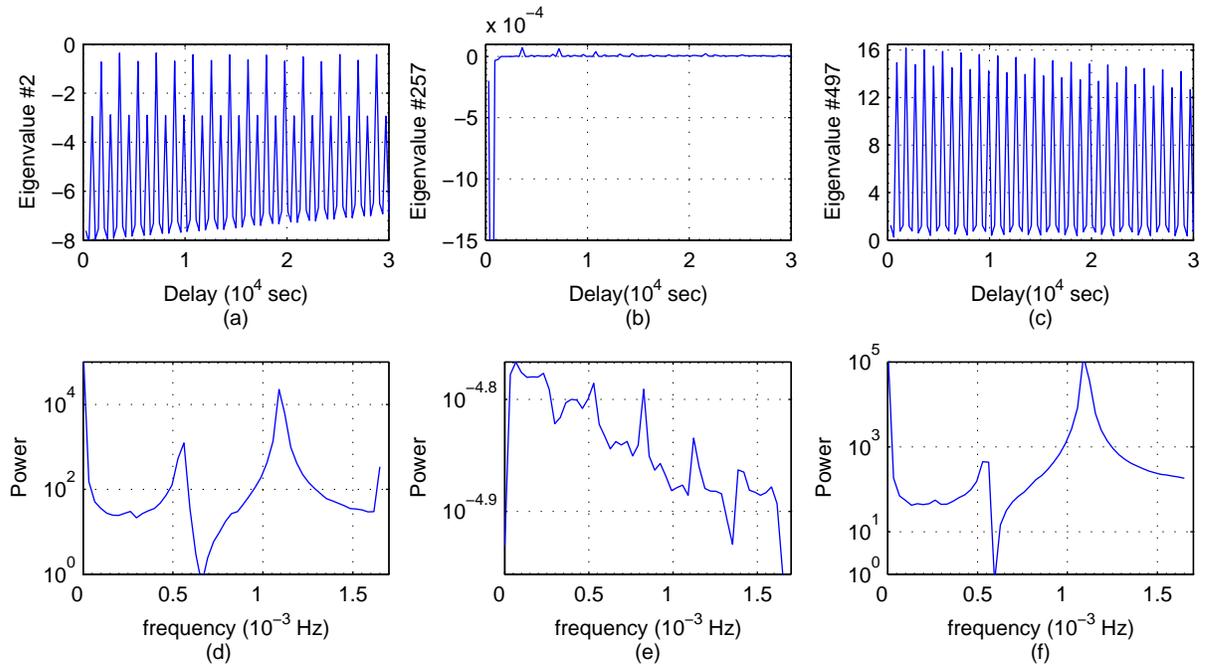}\end{center}

\caption{\label{7} Eigenvalues number $2$, $257$, and $497$: The results
of the injections with $2.5\, min$ period.}
\end{figure*}
\begin{figure*}
\begin{center}\includegraphics{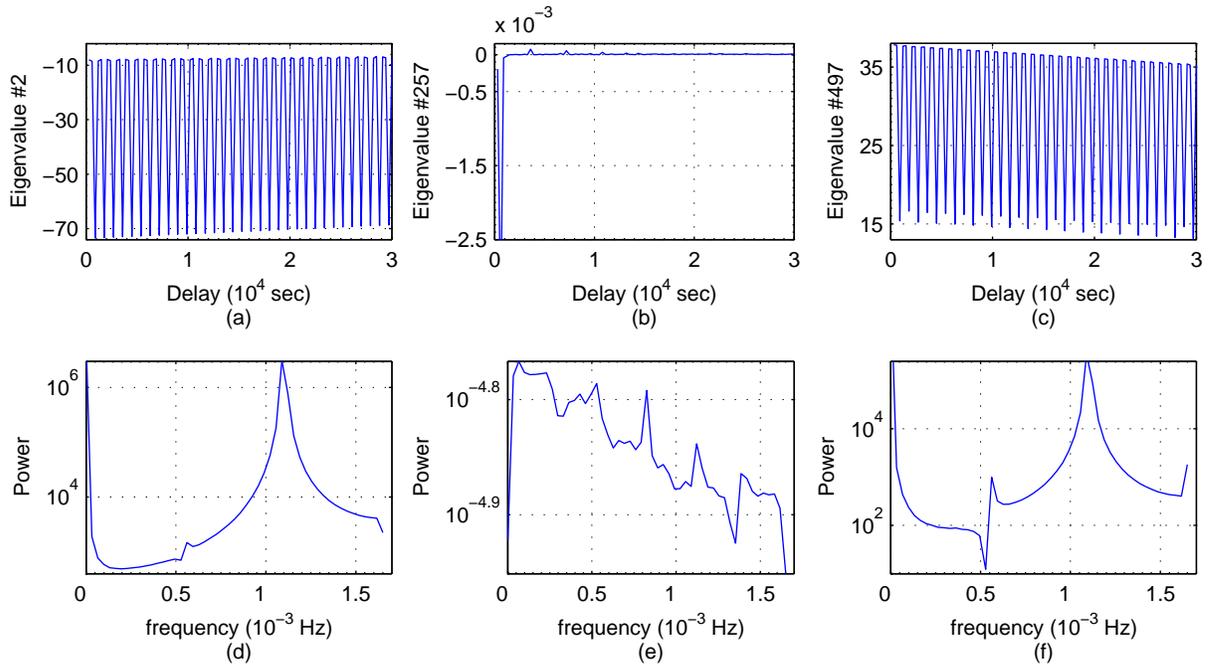}\end{center}

\caption{\label{8} Eigenvalues number $2$, $257$, and $497$: The results
of the injections with $15\, min$ period.}
\end{figure*}
A logical continuation of the above experiment is the injection of
an artificial traffic counts which possess regularity into actual
experimental data. This time, however, we perform the replacements
for the time series which can be traced back to the eigenvalues falling
into the random segment. Time series for this replacement were chosen
at random. Other possibilities can also be considered, but since random
segment was much less susceptible to the previous experiment, the
above choice seems natural.%
\begin{figure*}
\begin{center}\includegraphics{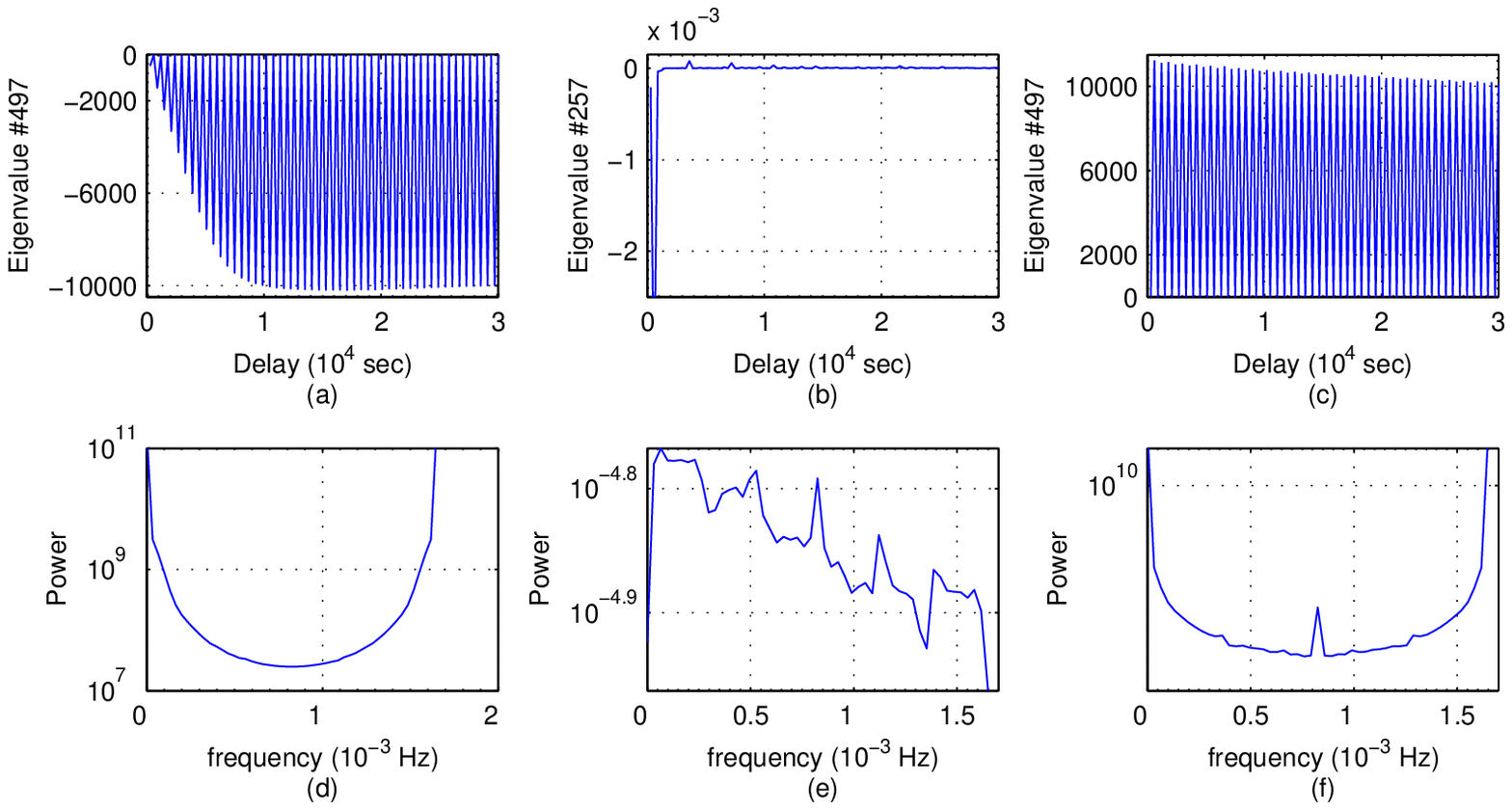}\end{center}

\caption{\label{9} Eigenvalues number $2$, $257$, and $497$: The results
of the injections with $20\, min$ period.}
\end{figure*}
\begin{figure*}
\begin{center}\includegraphics{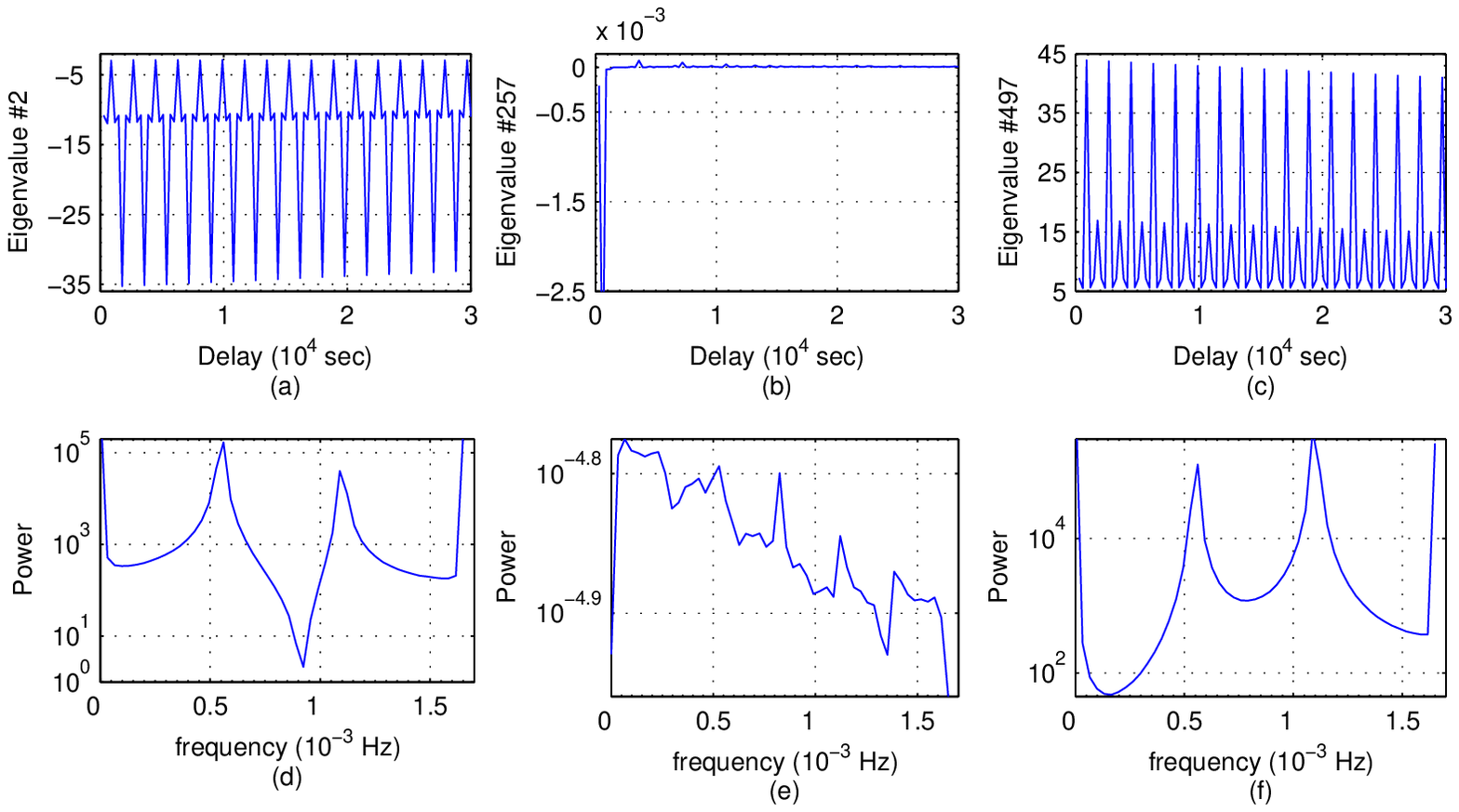}\end{center}

\caption{\label{10} Eigenvalues number $2$, $257$, and $497$: The results
of the injections with $30\, min$ period.}
\end{figure*}

We choose four injections to be cosinusoidal, having periods of $2.5$;
$15$; $20$ and $30\, min$ and repeating the same manipulations
as in the first experiment discussed in Section 4.1. The results are
fairly sound. Although the random part of the eigenvalue spectrum
is again unaltered, the {}``reaction'' of left and right parts is
both qualitative and quantitative. When cosinusoidal sample with period
much smaller than both characteristic periods ($15$ and $30\, min$),
power spectra in Figs. 7(d) and (f) are not significantly changed.
Two characteristic periods are still present, and yet certain narrow
frequency range gets suppressed (note anti-peak between the main two).
Notice slight asymmetry in the way smallest and largest eigenvalues
react to the injection. We should add, that we observed essentially
the same picture for the injections with period of $5$ and $10$
minutes (both not matching, but commensurate with characteristic periods).

Now, we turn to Fig. 8, where the cosinusoidal replacement with period
$15\, min$ of actual traffic counts leads to the dramatic change
in power spectrum. We observe enhancement of the peak corresponding
to period of $15\, min$, which can be termed as resonance phenomenon
(Figs. 8(d) and (f)). The very same plots show the suppression of
peaks corresponding to the other characteristic period of $30\, min$.
Similar resonant effect is achieved when the period of injection is
changed to $20\, min$ (see Fig. 9). This time, both peaks are gone,
while the new characteristic period is detected in Fig. 9(f). It is
approximately equal to the period of injection. Finally, for the experiment,
in which period of the injection was chosen to be $30\, min$, i.e.
matching to another characteristic period, we obtained yet another
result supporting previous conclusions.

In this case, however, the resonance phenomenon is slightly more difficult
to establish. From the results displayed in Fig. 10. We see, that
relative contribution to power spectrum is now changed for two main
peaks. Before the experiment was performed, the higher harmonic (smaller
period) dominated by quite a few orders of magnitude. After running
the experiment, this is still the case for the spectrum of largest
eigenvalue, but now the difference is marginal (see Fig. 10(d)). At
the same time for the left most eigenvalue, we determined, that lower
harmonic (period, matching the period of injection) now contributes
the most, as can be checked in Fig. 10(d). Two power spectra for the
edge eigenvalues are no longer symmetric and the contributions from
some ranges of frequencies are again strongly diminished. As for the
random eigenvalue considered Fig. 10(b) and (e), no impact has been
recorded, just as in all other cases.

\section{Discussion}

Long range or time dependent processes which show significant correlations
across large time scales were first discovered in network traffic
over a decade ago. Since then, LRD was found and studied intensively
in various aspects of network behavior. LRD is a manifestation of
self-similarity of the process, meaning that the behavior of the process
is space and time scale invariant. Leland and colleagues performed
first rigorous statistical analysis of self-similar characteristics
in Local Area Network (LAN) traffic \cite{Leland}. They showed that
the aggregated Ethernet traffic is not smoothing out with accordance
to Poisson model, it is time scale invariant. In this framework the
traditional Poisson or memory-less models of network traffic became
inadequate. Since high variability across different time scales produces
high congestion level, the impact of the self-similar traffic models
on queuing performance is considerable \cite{Erramilli}. 

However, the identification of self-similarity origin and estimation
of LRD in network traffic are far from straightforward. One of the
effective procedures to quantify LRD is to calculate the value of
Hurst parameter. Even though the Hurst parameter is a scalar it cannot
be calculated directly, it can be only estimated. There are several
methods to estimate the Hurst parameter and sometimes they produce
the conflicting results. Ineptly, it is not clear which method provides
the most accurate estimation \cite{Molnar,Krunz}. 

In this paper, we proposed the LRD and self-similarity indicators
of delayed correlations in network traffic. We demonstrated, that
the time delay invariant behavior of non-edge eigenvalues of $D\left(\tau\right)$
reflects the self-similar nature of delayed correlations. Meanwhile,
the scaling with time of edge eigenvalues or their lagged-time dependence
is an exhibition of self-similarity of delay correlations.

In addition, we established that the IPR for eigenvectors of lagged
correlations are concise parameters of realistic model for network
congestion pattern. As was shown in \cite{Rojkova}, IPR for $D\left(0\right)$
contains two localization trends in network interactions, i.e. two
regions in spectrum, traced back to a small number of time series,
which create the bottleneck at the routers. It is noteworthy, that
the IPR for $D\left(\tau\right),$ where $\tau>0$, reveals the third
localization trend, which has different origin. The significantly
increased and time delay invariant IPR around the median eigenvalue
indicates presence of lead-lag relationship between time series.

With experiments altering the original traffic time series several
distinctive effects has been uncovered. First of all, we demonstrated
that tempering with time series has no effect on self-similar transient
behavior of eigenvalues and IPRs, located in the middle segment of
the spectrum. By contrast, both stochastic and periodic injections
into the right (non-random) and middle (random) segments respectively
yielded dramatic changes in chosen indicators. In particular, we recorded
the destructive effect of random noise on otherwise simplistic double-peaked
power spectra.

One of the main results we obtained from periodic injection experiments
was presence of resonance phenomenon. When the period of injection
coincides with one of the characteristic time scales of the network
(i.e. oscillation periods of edge eigenvalues) the corresponding spectral
peak gets enhanced. The Fourier transform peak, corresponding to the
other scale gets suppressed and sometimes even annihilated. Finally,
injection with the period much less than both scales has little effect
on Fourier spectra, while period of the same order in magnitude rearranges
the original spectra completely.

The above described time-lagged correlational analysis has a broad
area of applications, where delayed correlations between system substructures
are essential. For instance, it can be applied to electro-physiological
time series of brain response \cite{Kwapien}, earthquake relocations
\cite{earthquake}, financial portfolios \cite{GlupyeIndusy,Biely},
and atmospheric data \cite{GlupyeIndusy}. To support this assertion
we point out that edge eigenvalues of $D\left(\tau\right)$ behave
almost identically to these of atmospheric data, while the delay eigenvalues
of the stock market data act just like the eigenvalues, we termed
random \cite{GlupyeIndusy}.

\section*{acknowledgment}

This research was partially supported by a grant from the US Department
of Treasury through a subcontract from the University of Kentucky.

\end{document}